\documentclass[traditabstract]{aa} 
\usepackage{graphicx}
\usepackage{txfonts}

\newcommand{\myrule}{\rule[-0.1cm]{0.cm}{0.5cm}} 

\newcommand\msun{M_{\odot}}

\newcommand\mjup{M_\mathrm{Jup}}

\newcommand\mperyr{M_{\odot}\,\mathrm{yr}^{-1}}
\newcommand\iso{ISO\,}

\begin{document}

\title{The bipolar outflow and disk of the brown dwarf \iso217\thanks{Based 
     on observations obtained at the Very Large Telescope of the 
	    European Southern Observatory at Paranal, Chile 
	    in program 
	    079.C-0375(A),    
	    080.C-0904(A), and   
	    082.C-0023(A+B).  
	  }}

 \titlerunning{The bipolar outflow and disk of the brown dwarf \iso217}

   \author{V. Joergens
          \inst{1,2}
          \and
          A. Pohl\inst{2}
	  \and
	  A. Sicilia-Aguilar\inst{3}
          \and
          Th. Henning\inst{2}
          }

   \institute{
	     Zentrum f\"ur Astronomie Heidelberg, 
	     Institut f\"ur Theoretische Astrophysik,
	     Albert-Ueberle-Str. 2, 69120 Heidelberg, Germany
	\and     
	     Max-Planck Institut f\"ur Astronomie, 
             K\"onigstuhl~17, D-69117 Heidelberg, Germany,
             \email{viki@mpia.de}
	 \and
	     Departamento de F{\'i}sica Te{\'o}rica,
	     Facultad de Ciencias,
	     Universidad Aut{\'o}noma de Madrid
	     Cantoblanco, 28049 Madrid, Spain
             }

   \date{Received on March 12, 2012; accepted on May, 6, 2012.}

  \abstract
   {We show that the very young brown dwarf candidate \iso217 (M6.25) is 
driving an intrinsically asymmetric bipolar outflow with a stronger and slightly faster red-shifted component based on spectro-astrometry 
of forbidden [S\,II] emission lines at 6716\,{\AA} and 6731\,{\AA} observed in UVES/VLT spectra taken in 2009.
\iso217 is only one of a handful of brown dwarfs and very low-mass stars 
(M5-M8) for which an outflow has been detected and that show that the T~Tauri phase continues
at the substellar limit.
We measure a spatial extension of the outflow in [S\,II] of up to $\pm$190\,mas (about $\pm$30\,AU) and
velocities of up to $\pm$40-50\,km\,s$^{-1}$. 
We find that the basic outflow properties (spatial extension, velocities, and outflow position angle) 
are of similar order as those determined in the discovery spectra from May 2007 of Whelan and coworkers.
We show that the velocity asymmetry between both lobes is variable on timescales of a few years
and that the strong asymmetry of a factor of two found in 2007
might be smaller than originally anticipated when using a more realistic stellar rest-velocity.
We also detect forbidden line emission of [Fe\,II]$\lambda$7155\,{\AA}, 
for which we propose as a potential origin the hot inner regions of the outflow.
To comprehensively understand the \iso217 system, 
we determine the properties of its accretion disk
based on radiative transfer modeling of the SED from 0.66 to 24\,$\mu$m. This disk model
agrees very well with Herschel/PACS data at 70\,$\mu$m.
We find that the disk is flared and intermediately inclined ($i\sim$45$^{\circ}$). 
The total disk mass of the best-fit model is 4$\times 10^{-6} M_\odot$,
which is low compared to the accretion and outflow rate of \iso217 from the literature ($\sim10^{-10}\,\mperyr$). 
We propose that this discrepancy can be explained by either a higher disk mass than inferred from the model because of
strong undetected grain growth and/or by an on average lower accretion rate and outflow rate
than the determined values.
We show that a disk inclination significantly exceeding 45$^{\circ}$, as 
suggested from H$\alpha$ modeling 
and from both lobes of the outflow being visible, 
is inconsistent with the SED data.
Thus, despite its intermediate inclination angle, 
the disk of this brown dwarf does not appear to obscure the red outflow component in [S\,II],
which is very rarely seen for T~Tauri objects (only one other case). 
}

\keywords{
		Stars: low-mass, brown dwarfs -
		Stars: pre-main sequence -
		Stars: circumstellar matter -
		Stars: formation -
		ISM: jets and outflows -
		Stars: individual (\mbox{\iso217})
} 

   \maketitle
%

\section{Introduction}
\label{sect:intro}

Jets and outflows are a by-product of accretion in the star formation process
(e.g., Ray et al. 2007 for a review).
They have been observed for many classical T~Tauri stars (CTTS)
in terms of the emission in atomic and molecular lines that originates
in the radiative cooling zones of shocks with moderate to large velocities (a few tens 
to a few hundred km\,s$^{-1}$).
These detections have been made either directly through narrow-band imaging,
e.g. in molecular lines of CO or forbidden emission
lines (FELs), 
or through spectro-astrometry of FELs (in some cases also of H$\alpha$ emission lines).
It has been suggested that jets transport a significant amount of excess angular-momentum
from the accretion disk, as some jets have been found to rotate (e.g., Launhardt et al. 2009).
The observed correlation between mass outflow and disk accretion 
indicates a magnetohydrodynamic jet-launching mechanism. 
The jet could originate from either a wide range of disk radii ('disk wind model'),
as favored by a high-resolution kinematic and collimation study of [Fe\,II] emission in DG~Tau
(Agra-Amboage et al. 2011),
or the interface between the star's magnetosphere and the disk ('X wind model').
Typical mass-loss rates for CTTS are found to be on the order of
$10^{-9}-10^{-7}\mperyr$, which is about 5\,\% to 10\,\% of the mass accretion rate
through the disk 
(e.g., Sicilia-Aguilar et al. 2006, 2010; Fang et al. 2009). 

The technique of spectro-astrometry plays a major role in 
probing the innermost region of jets, where the central engine is ($\lesssim$10\,AU)
and most of the collimation and acceleration occurs ($\lesssim$100\,AU),
and in detecting jets of very low-mass objects, for which the critical density for FELs
occurs very close to the driving source ($\sim$10-30\,AU).
Exploring jets on such small scales in nearby star-forming regions ($\sim$150\,pc)
requires milli-arcsecond (mas) resolution. 
Spectro-astrometry is a means of recovering spatial information well below the diffraction limit of the largest 
optical/near-infrared (IR) telescopes by
measuring the positional centroid of the emission as a function of wavelength
of an unresolved star in slit-spectroscopy. The spatial resolution depends here only on the 
ability to measure the centroid position.
The application of this method to CTTS was developed from pioneering work 
starting in the eighties (Solf 1984; Solf \& B\"ohm 1993; Hirth et al. 1994a, 1997) 
to resolve jets at 10\,AU from the central source
(e.g., Garcia et al. 1999, Takami et al. 2001, 2003) and to discover the jets of 
brown dwarfs (Whelan et al. 2005).

While many details about the origin of brown dwarfs are still unknown,
it has been established in the past few years that brown dwarfs during their early evolution
resemble higher mass T~Tauri stars in many properties. 
Very young brown dwarfs (a few Myr) display chromospheric activity, such as 
surface spots (e.g., Joergens et al. 2003).
There is evidence that brown dwarfs have disks from 
mid-IR (e.g., Comer\'on et al. 2000; Jayawardhana et al. 2003; Luhman et al. 2008) and
far-IR/submm excess emission (Klein et al. 2003; Scholz et al. 2006; 
Harvey et al. 2012a). Many of these disks have been found to 
be actively accreting (e.g., Mohanty et al. 2005; Herczeg \& Hillenbrand 2008; 
Bacciotti et al. 2011; Rigliaco et al. 2011) and
several show signs of grain growth and crystallization (e.g., Apai et al. 2005; Pascucci et al. 2009).
Furthermore, very young brown dwarfs rotate on average much slower 
(e.g., Joergens \& Guenther 2001; Joergens et al. 2003; Caballero et al. 2004) 
than their older counterparts (e.g., Bailer-Jones \& Mundt 2001; Mohanty \& Basri 2003), 
which is indicative of a magnetic braking mechanism due to interaction with the disk.

The first indication that brown dwarfs and very low-mass stars (VLMS) 
might also be able to drive T~Tauri-like outflows came from 
the observation of forbidden emission in the spectrum of an M6.5 dwarf, 
which is known to be an active accretor (LS-R\,CrA\,1, Fernandez \& Comer\'on 2001). 
Spectro-astrometry of detected forbidden [S\,II], [O\,I], and [NII] emission 
of several brown dwarfs and VLMS 
then provided proof that objects of a tenth of a solar mass to less than 30\,$\mjup$ 
can launch powerful outflows:
Par-Lup3-4 (M5), $\rho$\,Oph\,102 
(M5.5\footnote{Spectral type $\rho$\,Oph\,102: K. Luhman, private communication.}), 
\iso217 (M6.25), LS-R\,CrA\,1 (M6.5), 2M1207 (M8), and ISO-Oph\,32 (M8)
(Whelan et al. 2005, 2007, 2009a, 2009b;
Bacciotti et al. 2011; cf. also Fernandez \& Comer\'on 2005; Looper et al. 2010).
This was further supported by a resolved image of the outflow of the VLMS $\rho$\,Oph\,102 (M5.5),
which was detected in the CO J=2-1 transition 
with a 10\,arcsec spatial offset from the central source 
(Phan-Bao et al. 2008; cf. also Phan-Bao et al. 2011).
Directly resolving brown dwarf outflows in FELs is , on the other hand, challenging (e.g., Wang \& Henning 2006)
because the critical density of FELs occurs 
at very close separations (of $\sim$100\,mas at a distance of $\sim$150\,pc).

The investigated brown dwarfs and VLMS exhibiting outflows constitute a small sample of
six objects, out of which only four have a spectral type later than M6 
(\iso217, LS-R\,CrA\,1, 2M1207, ISO-Oph\,32), 
i.e. are at or below the substellar border.
They seem to have similar (scaled-down) properties as CTTS jets in several respects, as
for example they can be asymmetric and their FEL regions appear to contain both low and high velocity components.
First estimates of their mass-loss rates 
($\dot{M}_{out}=10^{-9}-10^{-10}\,\mperyr$, Whelan et al. 2009a; Bacciotti et al. 2011)
and comparisons with mass accretion rates for brown dwarfs 
($\dot{M}_{acc}=10^{-9}$ and $10^{-11}\,\mperyr$, 
e.g., Muzerolle et al. 2003; Natta et al. 2004; Mohanty et al. 2005; Herczeg \& Hillenbrand 2008)
give tentative hints of a relatively high $\dot{M}_{out}/\dot{M}_{acc}$ ratio, 
e.g. 40\,\% for Par-Lup3-4 (Bacciotti et al. 2011).
However, given the small number of only a handful of detected outflows for brown dwarfs and VLMS, 
for most of which only single epoch observations are made,
the available data do not yet provide a robust data set to help us establish their properties.
Important questions remain, concerning for example the degree of collimation and
possible variability. 

We explore here the bipolar outflow of the brown dwarf candidate \iso217 (M6.25)
by means of spectro-astrometry of high-resolution UVES/VLT spectra 
taken two years after the discovery data.
In a complementary fashion, we determine the disk properties of \iso217
based on radiative transfer modeling of its spectral energy distribution (SED) 
to comprehensively understand the disk and outflow system.
The paper is organized as follows: 
After a summary of the known properties of \iso217 (Sect.\,\ref{sect:iso217}),
the observations on which our work is based are described (Sect.\,\ref{sect:obs}). 
Section\,\ref{sect:disk} presents the modeling of the disk using flux measurements from the literature. 
In the next three sections, our high-resolution UVES spectra of \iso217 are exploited by studying emission line profiles
(Sect.\,\ref{sect:lines}), performing a spectro-astrometric analysis of forbidden [S\,II] emission 
(Sect.\,\ref{sect:spectroastro}), and analyzing our spectro-astrometric detection of the bipolar outflow
(Sect.\,\ref{sect:results}).
Section\,\ref{sect:concl} provides a discussion and conclusion of the \iso217 disk and outflow system.

\section{The brown dwarf candidate \iso217}
\label{sect:iso217}

\iso217\footnote{Simbad name: \object{ISO-ChaI 217}} is an M6.25 type very low-mass object 
(Muzerolle et al. 2005; Luhman 2007)
located in the Chamaeleon\,I (Cha\,I) star-forming region at a distance of $\sim$160-165\,pc. 
An estimate of its mass based on a comparison of effective temperature and luminosity
($T_\mathrm{eff}$=2962\,K, $L_\mathrm{bol}$=0.023\,$L_{\odot}$, Luhman 2007)
with evolutionary models (Baraffe et al. 1998)
yields a value of about 0.08\,$\msun$, i.e. close to the hydrogen burning limit.

Mid-IR excess emission of \iso217 was detected by the {\em Infrared Space Observatory Camera} (ISOCAM, 
Persi et al. 2000; Lopez Marti et al. 2004)
and the Spitzer space mission (Apai et al. 2005; Pascucci et al. 2009; Luhman et al. 2008) 
indicating that \iso217 has a disk.
A strong 10\,$\mu$m silicate emission detected in a spectrum taken by the 
Spitzer/{\em InfraRed Spectrograph (IRS) gives 
evidence of both grain growth and moderate crystallization in this disk  
(Apai et al. 2005).} 
The disk inclination was suggested to be 65$^{\circ}$ 
based on modeling of the H$\alpha$ line profile (Muzerolle et al. 2005;
the inclination is defined here as the angle between the line-of-sight and the stellar rotation axis, 
so that 90$^{\circ}$ corresponds to an edge-on system, J. Muzerolle, private communication).

\iso217 displays broad H$\alpha$ emission with an equivalent width (EW) varying 
between 70\,{\AA} and 230\,{\AA} indicating ongoing variable disk accretion
(Muzerolle et al. 2005; Luhman 2004; Scholz \& Jayawardhana 2006; this work). 
Scholz \& Jayawardhana (2006) found that the variability of the H$\alpha$ line can be predominantly attributed to
the emission wings and thus to high-velocity infalling gas.
Furthermore, a blue-shifted absorption dip present in the H$\alpha$ profile 
(Scholz \& Jayawardhana 2006; Whelan et al. 2009a) appears to correspond to 
a wind with a velocity of between a few and about 30\,km\,s$^{-1}$.
An estimate of the accretion rate based on a magnetospheric model of the H$\alpha$ line profile at one epoch 
yields 1.0$\times$10$^{-10}\,\mperyr$ (Muzerolle et al. 2005), 
which is a typical value for brown dwarfs and VLMS.

The presence of forbidden [S\,II]$\lambda$6731 line emission in some spectra of \iso217
taken by Scholz \& Jayawardhana (2006) gave the first hints for the 
outflow activity of this very low-mass object.
A spectro-astrometric analysis of forbidden line emission in 
[S\,II]$\lambda$6731, [S\,II]$\lambda$6716, [O\,I]$\lambda$6300, and [O\,I]$\lambda$6363
revealed that \iso217 is driving a bipolar outflow 
with an estimated mass-loss rate of 2-3$\times10^{-10}\,\mperyr$
(Whelan et al. 2009a).
We explore the properties of this bipolar outflow
based on spectro-astrometry of [S\,II]$\lambda$6731 and [S\,II]$\lambda$6716 emission recorded 
two years after the discovery data. Furthermore, 
we determine the disk properties of \iso217
by means of SED modeling.

\section{Observations}
\label{sect:obs}

\subsection{The spectral energy distribution of \iso217} 
\label{sect:obs1}

To model the disk of \iso217 (Sect.\,\ref{sect:disk}), 
we used optical to mid-IR flux measurements from the literature.
The existing photometry in optical (RI) and near-IR (JHK) bands is mainly based on 
observations by the ESO 2.2m/{\em Wide Field Imager} (Lopez Marti et al. 2004),
the {\em Two Micron All Sky Survey}, and the 
{\em Deep Near Infrared Survey of the Southern Sky} (e.g., Carpenter et al. 2002).
Mid-IR photometry of \iso217 was obtained by ISOCAM (6.7\,$\mu$m, Persi et al. 2000; Lopez Marti et al. 2004),
the {\em InfraRed Array Camera} (IRAC) on board the Spitzer satellite (3.6, 4.5, 5.8, 8.0\,$\mu$m), 
and the {\em Multiband Imaging Photometer for Spitzer} (MIPS, 24\,$\mu$m, Luhman et al. 2008). 
In addition, a Spitzer/IRS spectrum ($7.4-14.5\,\mu$m) was taken (Apai et al. 2005; Pascucci et al. 2009).
Very recently, \iso217 was observed in the far-IR by  the
{\em Photoconductor Array Camera and Spectrometer} (PACS) of the Herschel mission
(Harvey et al. 2012b). We show in Sect.\,\ref{sect:disk} that our disk model, 
which is based on the SED up to 24\,$\mu$m,
is in very good agreement with the Herschel flux measurement at 70\,$\mu$m.

{\em Photometric variability.}
The star appears to be significantly variable at optical and IR wavelengths (Fig.\,\ref{fig:disk}).
In particular, the IRS observations indicate that the flux is higher by nearly a factor of two
than in the IRAC 8\,$\mu$m observation. The ISOCAM data, on the other hand,
detect a flux that is about 30\% fainter than the IRAC data. 
This strong variability from optical to mid-IR wavelengths 
suggests that the cause of the variations is an increase in the luminosity of the star 
(for instance, owing to a higher accretion rate) rather than an occultation effect.
On the basis of the observed variability in the 8\,$\mu$m region, it 
can be expected that the 24\,$\mu$m flux is also variable for this object
(e.g., Muzerolle et al. 2009).  
The single-epoch MIPS data were not taken simultaneously with any other photometry,  
therefore the average 24\,$\mu$m flux level and the 8-24\,$\mu$m slope
remain somewhat uncertain. This as well as 
the the mid-IR data not being taken simultaneously with the optical/near-IR data, which represent the (sub)stellar 
photosphere, complicated the SED modeling (Sect.\,\ref{sect:disk}).

\subsection{High-resolution spectroscopy of \iso217}
\label{sect:obs2}

\begin{table*}
\begin{minipage}[b]{\columnwidth}
\begin{center}
\caption{
\label{tab:obslog} 
Observing log, slit position angle, and radial velocity of \iso217. 
}
\renewcommand{\footnoterule}{}  
\begin{tabular}{lllllll}
\hline
\hline
\myrule
Name & Date   & HJD           & EXPTIME & Seeing   & Slit PA    & RV          \\
     &        &               & [sec]   & [arcsec] & [deg] & [km\,s$^{-1}$] \\
\hline
\myrule
spec\,1 & 2008 03 22 & 2454547.61149 & 2$\times$1500 & 0.93 & 158.9\,$\pm$\,7.3 \\
spec\,2 & 2008 03 22 & 2454547.64793 & 2$\times$1500 & 0.78 & 173.8\,$\pm$\,7.4 & 16.8\,$\pm$\,2.2\\
\hline
\myrule
spec\,3 & 2009 01 30 & 2454861.68128 & 2$\times$1500 & 1.08 & 131.7\,$\pm$\,6.9 \\
spec\,4 & 2009 01 30 & 2454861.71766 & 2$\times$1500 & 1.10 & 146.0\,$\pm$\,7.1 & 17.4\,$\pm$\,1.6\\
\hline
\myrule
spec\,5 & 2009 02 23 & 2454885.70518 & 2$\times$1500 & 0.79 & 167.2\,$\pm$\,7.4 \\
spec\,6 & 2009 02 23 & 2454885.74161 & 2$\times$1500 & 0.69 & 182.3\,$\pm$\,7.5 & 17.4\,$\pm$\,2.4\\
\hline
\hline
\end{tabular}
\tablefoot{
The HJD is given at the middle of the exposure; 
the seeing is the averaged seeing corrected by airmass; 
the slit position angle (PA) was not kept fixed during observations
and each 2$\times$1500\,sec exposure samples a 
PA range of about $\pm$\,7\,deg, as listed; 
the radial velocity (RV) is determined based on a fit to the Li\,I ($\lambda$ 6708\,\AA) line
in the averaged spectra of each night.
}
\end{center}
\end{minipage}
\end{table*}

Spectroscopic observations of \iso217 were obtained
within the framework of a high-resolution
spectroscopic study of young brown dwarfs and VLMS in Cha\,I 
(e.g., Joergens 2006, 2008). 
\iso217 was observed for three nights in 2008 and 2009 in the red optical-wavelength regime with
the \emph{Ultraviolet and Visual Echelle Spectrograph} (UVES, Dekker et al. 2000) 
attached to the VLT 8.2\,m KUEYEN telescope.
The spectral resolution $\lambda$/$\Delta \lambda$ was 40\,000 and
the spatial sampling 0.182$^{\prime\prime}$/pixel.
An observing log is given in Table\,\ref{tab:obslog}.
We used UVES spectra that were reduced, and both wavelength- and flux-calibrated using the ESO UVES pipeline
in order to study activity-related emission lines (Ca\,II, [S\,II], [Fe\,II]) in terms of
their line profile shapes and line strengths (Sect.\,\ref{sect:lines}).
Furthermore, a spectro-astrometric analysis of the detected forbidden line
emission in [S\,II] was performed based on a custom-made reduction and wavelength calibration
procedure (Sect.\,\ref{sect:spectroastro}).

{\em Stellar rest-velocity.}
The stellar rest-velocity of \iso217 was determined based on a Gaussian fit to 
the Li\,I doublet lines at 6707.76\,\r{A} and 6707.91\,\r{A} in the pipeline-reduced spectra.
This Li\,I doublet was not resolved in our observations, therefore, their mean value (6707.84\,\r{A}) was 
adopted as a reference in the velocity measurement. 
We note that the maximum error possibly introduced by the assumption of two equally strong lines was 
estimated to be less than 
1\,km\,s$^{-1}$ by using a high-resolution synthetic PHOENIX spectrum of similar stellar parameters
(Husser et al. 2012).
The radial velocities determined in this way 
and their errors, which are based on the Gaussian fit, are listed in Table\,\ref{tab:obslog} for the 
average spectra of each 
night.
They agree well with the radial velocities measured for other 
brown dwarfs and T~Tauri stars in Cha\,I (e.g. Covino et al. 1997; Joergens 2006; James et al. 2006).
Their weighted mean value $V_{0}=(17.2\pm1.3)$\,km\,s$^{-1}$ was adopted as 
stellar rest-velocity for \iso217, relative to which all velocities in the following are given.

\section{Modeling the disk of \iso217}
\label{sect:disk}

\begin{figure}[b]
\centering
\includegraphics[width=.95\linewidth,clip]{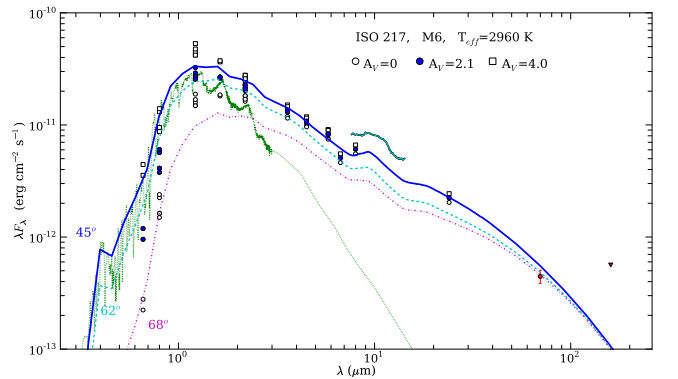}
\caption{
\label{fig:disk}
{\bf SED and disk model for \iso217.} 
Flux measurements are plotted for extinction values of Av=0, 2.1, and 4\,mag, as denoted.
The displayed models are a PHOENIX photosphere model (green dotted line) 
and our best-fit disk model for an inclination of $i$=45$^{\circ}$ (dark blue line).
For comparison, we also show disk models for $i$=62$^{\circ}$ (turquoise dashed line)
and 68$^{\circ}$ (pink dash-dotted line).
}
\end{figure}

The outflow activity of young stellar objects is intrinsically tied to the accretion disk.
To gain a comprehensive understanding of the \iso217 outflow and disk system,
we determined the disk properties (structure, orientation, and mass) of \iso217
by means of radiative transfer modeling of its SED (Fig.\,\ref{fig:disk}).
A disk has been detected around \iso217 based on mid-IR excess emission
(cf. Sect.\,\ref{sect:iso217} and \ref{sect:obs1}), though no SED modeling of this disk has yet 
been performed.
The suggested high inclination-angle of the \iso217 disk, which was inferred
from a modeling of the H$\alpha$ profile (65$^{\circ}$, Muzerolle et al. 2005)
and from both lobes of the outflow being visible (Whelan et al. 2009a),
requires verification by an SED model. 

We fitted the observed SED of \iso217 from the optical to the mid-IR 
(0.66 - 24\,$\mu$m, cf. Sect.\,\ref{sect:obs1})
with the radiative transfer code RADMC\footnote{See http://www.ita.uni-heidelberg.de/~dullemond/software/radmc/}
of Dullemond \& Dominik (2004). 
The RADMC package was designed to model three-dimensional (3-D) axisymmetric 
circumstellar-dust configurations, and was tested extensively for continuum radiative transfer
in protoplanetary disks. Although assuming axisymmetry reduces
the problem to two dimensions, the photon packages are followed in 3-D.
The code uses a variant of the Monte Carlo method of Bjorkman
\& Wood (2001) to compute how the stellar photons
penetrate the disk and to determine the dust temperature and scattering
source function at any location in the disk. 
Applying volume ray-tracing allows us to determine 
the spectra and images for all inclination angles.
The code can be modified by selecting disk mass, flaring, outer
radius, vertical height of the disk at the outer
radius, and dust grain distribution. 
In all cases, it is assumed that the dust and gas are well-mixed (i.e. there is no differential
settling for grains with different masses) 
and that all grains have the same temperature distribution. 

For the photosphere of the star, we used a PHOENIX model (Hauschildt et al. 1999) 
with $T_{\rm{eff}}$=2960\,K and log(g)=4.0 (green dotted line in Fig.\,\ref{fig:disk}). 
To reproduce the integrated luminosity, the stellar radius was set to $R_*$=0.6\,R$_\odot$.
Since there is no evidence of an inner disk hole,
we located the inner rim of the disk at the dust destruction radius,
which for a dust destruction temperature of 1500\,K
is about 4\,R$_*$.
This agrees with the corotation radius (3--9\,R$_*$)
that was derived for \iso217 by considering typical rotational periods
of young brown dwarfs and VLMS 
(1-5\,d, Joergens et al. 2003; Rodr\'iguez-Ledesma et al. 2009).
The outer disk radius was chosen to be $R_{disk}$=100\,AU, 
although it is not well-constrained by any data. 
The disk mass was varied and then determined by requesting a good fit to the 24\,$\mu$m data point.

The object was assumed to have a moderate extinction, therefore we
considered A$_V$ values between 0\,mag and 4\,mag and a standard extinction law. 
Different A$_V$ values do not have a strong effect on the IR data, but can significantly 
change the optical fluxes.
For our best-fit model (dark blue line in Fig.\,\ref{fig:disk}), 
we adopted the spectroscopically measured extinction of A$_V$=2.1\,mag 
(converted from $A_J$=0.68, Luhman 2007, using 
the reddening law of Mathis 1990 and $R_V$=5.0, Luhman 2004). 

The strong variability seen for ISO\,217 in the optical to the 8\,$\mu$m region 
is a major challenge for SED modeling, in particular given that the data points were not obtained simultaneously
(cf. Sect.\,\ref{sect:obs1}). 
It is indicative of variations in both the accretion rate and luminosity and 
can be expected to affect also the 24\,$\mu$m flux (e.g., Muzerolle et al. 2009)
and the silicate feature (Abrah\'{a}m et al. 2009; Juh\'{a}sz et al. 2012).
Fortunately, \iso217 was observed with Herschel/PACS at 70\,$\mu$m (Harvey et al. 2012b), 
i.e. at wavelengths that play a crucial role in defining the structure and extent of
disk material and that are much less affected by variability.
We later show that the 70\,$\mu$m flux measurement strongly supports our approach to the SED fitting.
We traced an approximate model focused on the intermediate optical fluxes
and the IRAC and MIPS data. 
Since the IRAC data appear in-between the IRS and 
ISOCAM observations, we also considered intermediate
optical fluxes to estimate the stellar luminosity. We concentrated
on reproducing the SED slope between 3\,$\mu$m and 24\,$\mu$m and on
obtaining a silicate feature similar to the observed one in the IRS spectrum. 

The high flux level of \iso217 in the near-IR suggests that the
disk is very flared in its inner regions.
The relatively low flux level at 24\,$\mu$m, on the other hand,
could be reproduced by a model that has either a small amount of few-micron-sized grains and/or a  
low disk flaring and/or a low dust mass.
Since the near-IR fluxes indicate a very flared disk
and the silicate feature the presence of small grains (e.g., Henning 2010),
the low 24\,$\mu$m flux is most plausibly explained by a low dust mass.
We therefore applied a standard flaring law with a pressure scale-height (H$_p$) 
that varies as a power law with the radius, H$_p$/R$\propto$R$^{1/7}$.
The best-fit scale-height at the outer disk radius was found to be H$_{rdisk}$/R$_{disk}$=0.35. 

The dusty disk component in the model consists of amorphous Mg-Fe silicates
(with Fe and Mg being present in similar proportions; J\"ager et al. 1994; Dorschner et
al. 1995) and 25\% of amorphous carbon grains.
While there are signs of crystallization in the 
disk of \iso217 from the 10\,$\mu$m silicate feature (Apai et al. 2005),   
we found that the energy emitted in crystalline features is negligible compared to that from 
the continuum plus amorphous features and, thus, that crystallization
is not relevant for the SED modeling.
We considered a standard dust distribution with grain sizes of between 0.1\,$\mu$m
and 100\,$\mu$m, that follows a collisional distribution (power law)
with an exponent -3.5 (for both silicate and carbon grains alike).
Furthermore, a standard gas to dust ratio of 100 was assumed.
The best fit was achieved for a model with
a total disk mass of 4$\times 10^{-6} M_\odot$.

The exploration of different disk orientations yields a best-fit for
an intermediate inclination angle $i$ of 45$^{\circ}$
(dark blue line in Fig.\,\ref{fig:disk}).
We found that a disk with the described characteristics that has 
an inclination significantly exceeding 45$^{\circ}$, 
as suggested for example by H$\alpha$ modeling (65$^{\circ}$, Muzerolle et al. 2005),
is hard to explain: 
a disk with an inclination in the range 60$^{\circ}$-70$^{\circ}$
(see Fig.\,\ref{fig:disk} for models with $i$=62$^{\circ}$ and 68$^{\circ}$)
would not only be inconsistent with the near-IR data, but also produce a much higher extinction than 
the spectroscopically determined value and
compromise the presence of the relatively strong silicate feature.
Furthermore, a high extinction would require a flat, settled disk as well as 
a significantly lower disk mass in order to fit the steep slope at 3-24\,$\mu$m. 
A very settled disk, however, is unlikely because high levels of turbulence
can be expected given that
the object is accreting at a high rate for its
low mass and for its low disk mass.
Considering that the 24\,$\mu$m flux might be variable 
and measured at minimum, only the assumption of a very settled disk
would allow for a higher inclination.
To summarize, an inclination angle exceeding 60$^{\circ}$ is hard
to fit within a reasonable disk model.

Our disk model, which is based on flux measurements up to 24\,$\mu$m, is 
in very good agreement (within ~1.5\,$\sigma$) with recent Herschel/PACS 
observations of \iso217 at 70\,$\mu$m (Harvey et al. 2012b), as shown in Fig.\,\ref{fig:disk}.
This gives us confidence in 
the disk model developed  here 
because at these longer wavelengths the dependence on the mass and the grain
distribution is much stronger, while variability is expected to play a minor role.

The disk mass of \iso217 derived with our model 
of 4$\times 10^{-6} M_\odot$, i.e. about 1 Earth mass, is
very low for a CTTS disk but fully consistent with that of other brown dwarfs and VLMS 
(10$^{-5} - 10^{-6} M_\odot$, e.g., Harvey et al. 2012a).
However, it is strikingly low compared to the estimated accretion
($1\times10^{-10}\,\mperyr$, Muzerolle et al. 2005) 
and mass-loss rate (2-3$\times10^{-10}\,\mperyr$, Whelan et al. 2009a) of \iso217. 
Taking these numbers at face value, the total disk mass 
would be accreted and lost again within less than 40\,000\,yr, which is unlikely. 
These discrepancies between dust-inferred disk masses and gas-inferred mass accretion rates
are frequently found for CTTS (Hartmann 2008; Sicilia-Aguilar et al. 2011) 
and are usually explained in terms of a strong grain growth and/or anomalous gas-to-dust ratio.
The available SED data for \iso217 are only sensitive to the
presence of small grains because
particles with much larger sizes than the longest PACS wavelength do 
not contribute significantly to the flux owing to their small opacities. 
Considering a population of large ($>$100 $\mu$m) grains 
would result in a higher disk mass 
and might, therefore, account for some 
of the discrepancy with the gas-inferred mass-accretion rate.
However, a disk mass that allows for accretion 
over a typical disk lifetime of 10$^6$\,yr, would 
require a hundred times more mass to exist in these large grains.
Another possibility is that the accretion rate of \iso217 was measured in a 
high state and that it is on average lower 
than the determined value. This is plausible since
this value is based on single epoch observations and since
there are several indications of variable accretion 
of \iso217 from photometric (Sect.\,\ref{sect:obs1}) and emission-line variability 
(Scholz \& Jayawardhana 2006, Luhman 2007). 
In this case, there would still remain a mismatch with the outflow rate,
which might be a hint that it is 
also on average lower than the determined value.

The disk around \iso217 appears at first glance to be in a rather early evolutionary phase 
given its flared geometry, however, the possibility of undetected strong grain-growth
might hint at a more advanced evolutionary stage.
\iso217's disk is consistent with a flared, CTTS-like disk, 
in agreement with models for many other brown dwarf and VLMS disks 
(e.g., Natta \& Testi 2001; Allers et al. 2006).
Several flattened and evolved disks have been
identified around objects with similar masses (e.g. Pascucci et al.
2003; Morrow et al. 2008), and are usually traced to more evolved
systems with strong settling and grain growth. From this point of view,
the disk around \iso217 could be interpreted as a system that has
suffered little evolution. 
On the other hand, given the possibility of undetected 
large grains, as described in the previous paragraph, \iso217 is a good candidate 
for strong grain growth, despite the flared and primordial appearance of its disk.


\section{Observed emission lines of \iso217}
\label{sect:lines}

\begin{figure}[b]
\centering
\includegraphics[width=.95\linewidth,clip]{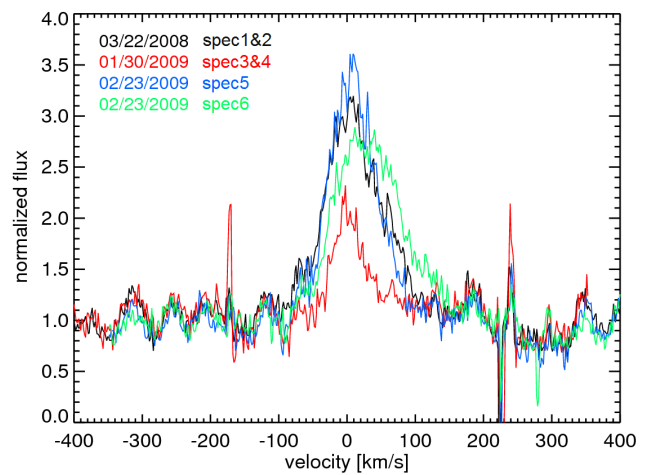}
\caption{
\label{fig:caII}
{\bf Ca\,II$\lambda$8498 emission line} in UVES spectra of \iso217
at different observing times in 2008 and 2009. 
All fluxes have been normalized to the continuum. 
}
\end{figure}

Our UVES spectra of \iso217 show several emission lines,
which are indicative of ongoing accretion, winds, and/or outflowing material
(Ca\,II, [S\,II], [Fe\,II]).
We studied their line profile shapes and measured their EWs.
For this purpose, we used the UVES spectra (Table\,\ref{tab:obslog})
after reduction, wavelength and flux calibration by the ESO UVES pipeline. 
Furthermore, the spectral regions of the emission lines were normalized by dividing
by a polynomial fit to the continuum emission adjacent to each line.
Line profile variations were found to be negligible within one observational night in most cases,
allowing us to average spectra from the same night to increase the signal-to-noise ratio (S/R).
There was one exception from this, namely the Ca\,II line profile in the third night
(Feb 23, 2009), which changed significantly on timescales of hours (Fig.\,\ref{fig:caII}) and was, 
therefore, analyzed in the individual spectra.
Table\,\ref{tab:ew} lists all observed emission lines, their peak wavelength $\lambda$,
peak radial velocities $v$, and the measured EW.
The error in the EW is assumed to be 5\% of the determined value.
The peak values ($\lambda$, $v$) were determined  
by fitting the line profiles with Gaussian functions and their errors 
were based on the errors in the fit parameters. In the case of the two lobes of the [S\,II] lines
as well as the [Fe\,II] line,
these errors were underestimated because of deviations from a Gaussian shape.
A detailed description of the results for each emission line is provided in the following.

\begin{figure}[b]
\centering
\includegraphics[width=.95\linewidth,clip]{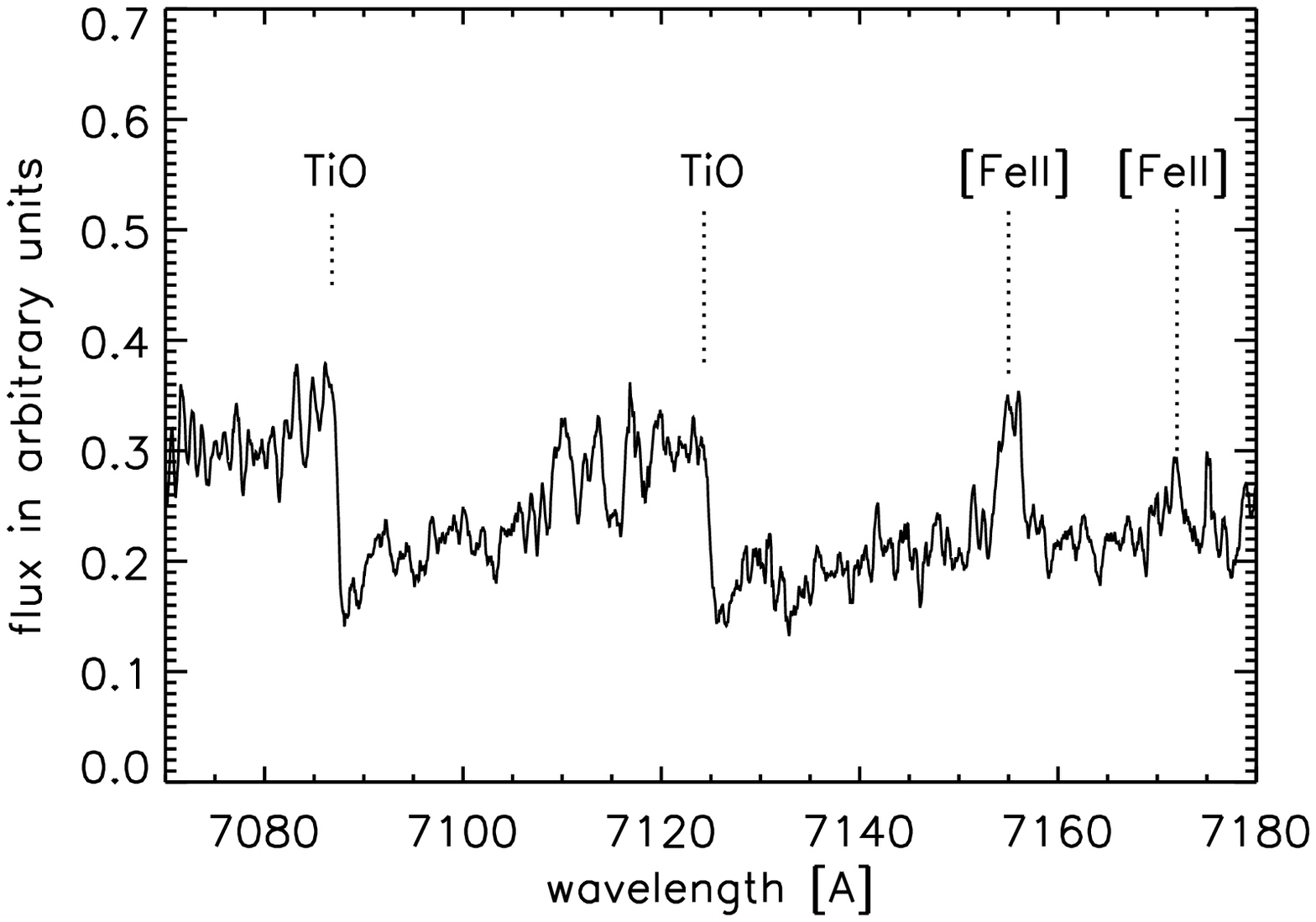}
\caption{
\label{fig:feII}
{\bf [Fe\,II] lines at 7155\,{\AA} and at 7172\,{\AA}} in UVES spectrum of \iso217.
Also displayed are two TiO absorption band-heads at 7087\,{\AA} and 7124\,{\AA} for comparison.
}
\end{figure}

{\em Ca\,II IR emission.}
\iso217 has a broad, asymmetric and variable Ca\,II IR emission line at 8498\,{\AA} (Fig.\,\ref{fig:caII}),
which implies that this line originates not purely in the chromosphere but also from accretion and/or winds.
We note that the other two lines of this IR triplet 
are not covered by the observations because of a gap in wavelength between the chips of the two-armed spectrograph.
The asymmetric Ca\,II profile displays a red-shifted large velocity tail with
velocities of up to $130\,$km\,s$^{-1}$, which could be caused by infalling material.
The shape, the peak flux, and the peak velocity of this line vary significantly for spectra 
taken at different epochs. 
The line profile shape observed in the third night (Feb 23, 2009;
green and blue profiles in Fig.\,\ref{fig:caII}) 
changed remarkably on timescales of hours.
The EW of this line do not generally follow these variabilities; we measured a
relatively constant value of about -6\,{\AA} apart from the second night (Jan 30, 2009), where it was 
-2\,{\AA}.

\begin{table*}
\begin{minipage}[t]{\columnwidth}
\begin{center}
\caption{
\label{tab:ew} 
Observed emission lines of \iso217.
}
\renewcommand{\footnoterule}{}  

\begin{tabular}{clcccc}
		
\hline\hline  
\myrule
& date & lobe & \multicolumn{2}{c}{peak} & \multicolumn{1}{c}{EW}  \\[0.15cm]
& & & \multicolumn{1}{c}{$\lambda$ [\r{A}]} & $v$ [km/s] & \multicolumn{1}{c}{[\r{A}]} \\[0.15cm]

\hline 
\myrule
Ca\,II & 2008 03 22  & & 8498.45$\pm$0.02 & 8.4$\pm$1.4 & -6.2$\pm$0.3 \\
(8498.02\,{\AA}) & 2009 01 30  & & 8498.22$\pm$0.03 & 1.8$\pm$1.5 & -2.0$\pm$0.1 \\
& 2009 02 23  (spec\,5) & & 8498.39$\pm$0.02 & 8.1$\pm$1.3 & -6.1$\pm$0.3 \\
& 2009 02 23  (spec\,6) & & 8499.01$\pm$0.02 & 29.8$\pm$1.4 & -6.3$\pm$0.3 \\
& 2009 02 23  (spec5$\&6$) & & 8498.63$\pm$0.02 & 16.4$\pm$1.4 & -6.2$\pm$0.3 \\[0.15cm]
\hline 
\myrule

[S\,II] & 2008 03 22  & red & 6731.75$\pm$0.01 & 34.2$\pm$1.2 & -4.5$\pm$0.2  \\
(6730.82\,{\AA}) & 2008 03 22  & blue & 6730.46$\pm$0.01 & -23.0$\pm$1.2 &  \\
& 2009 01 30  & red & 6731.69$\pm$0.01 & 33.6$\pm$1.2 & -5.3$\pm$0.3 \\
& 2009 01 30  & blue & 6730.36$\pm$0.01 & -25.7$\pm$1.2 &   \\
& 2009 02 23  & red & 6731.71$\pm$0.01 & 34.8$\pm$1.2 & -4.3$\pm$0.2  \\
& 2009 02 23  & blue & 6730.54$\pm$0.01 & -17.4$\pm$1.2 &   \\[0.15cm]

\hline 
\myrule
[S\,II] & 2008 03 22  & red & 6717.35$\pm$0.01 & 33.4$\pm$1.2 & -2.5$\pm$0.1  \\
(6716.44\,{\AA}) & 2008 03 22  & blue & 6716.28$\pm$0.02 & -14.1$\pm$1.4 &   \\
& 2009 01 30  & red &  6717.33$\pm$0.01 & 34.4$\pm$1.2 &  -3.4$\pm$0.2 \\
& 2009 01 30  & blue & 6716.17$\pm$0.02 & -17.5$\pm$1.4 &   \\	
& 2009 02 23  & red & 6717.34$\pm$0.01 & 35.0$\pm$1.2 & -2.4$\pm$0.1   \\
& 2009 02 23  &  blue & 6716.16$\pm$0.01 & -17.6$\pm$1.2 &   \\[0.15cm]

\hline 
\myrule
[Fe\,II]         & 2009 02 23 & & 7155.21$\pm$0.05 & -2.89$\pm$2.1 & 1.6$\pm$0.1 \\ 
(7155.16\,{\AA}) & & & \\

\hline
\hline
\end{tabular}
\tablefoot{The laboratory wavelengths of the emission lines are taken from the 
NIST database (http://physics.nist.gov/asd3, Ralchenko et al. 2011).}
The error in the EW is assumed to be 5\% of the determined value.
The errors in the peak values ($\lambda$, $v$) 
are based on the errors in the Gaussian fit parameters. 
The errors in $\lambda$ and $v$ are underestimated
in the case of deviations from a Gaussian shape ([S\,II], [Fe\,II]).
See the text for more details.
\end{center}
\end{minipage}
\end{table*}

\begin{figure}[t]
\centering
\includegraphics[width=.8\linewidth,clip]{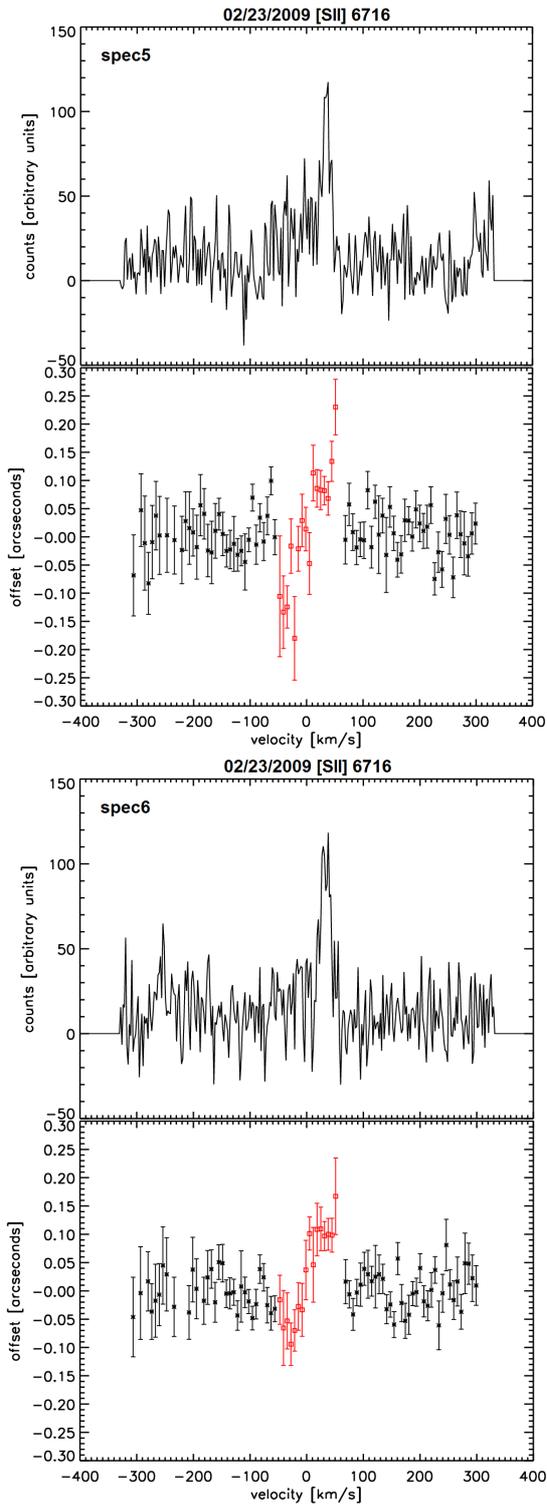}
\includegraphics[width=.8\linewidth,clip]{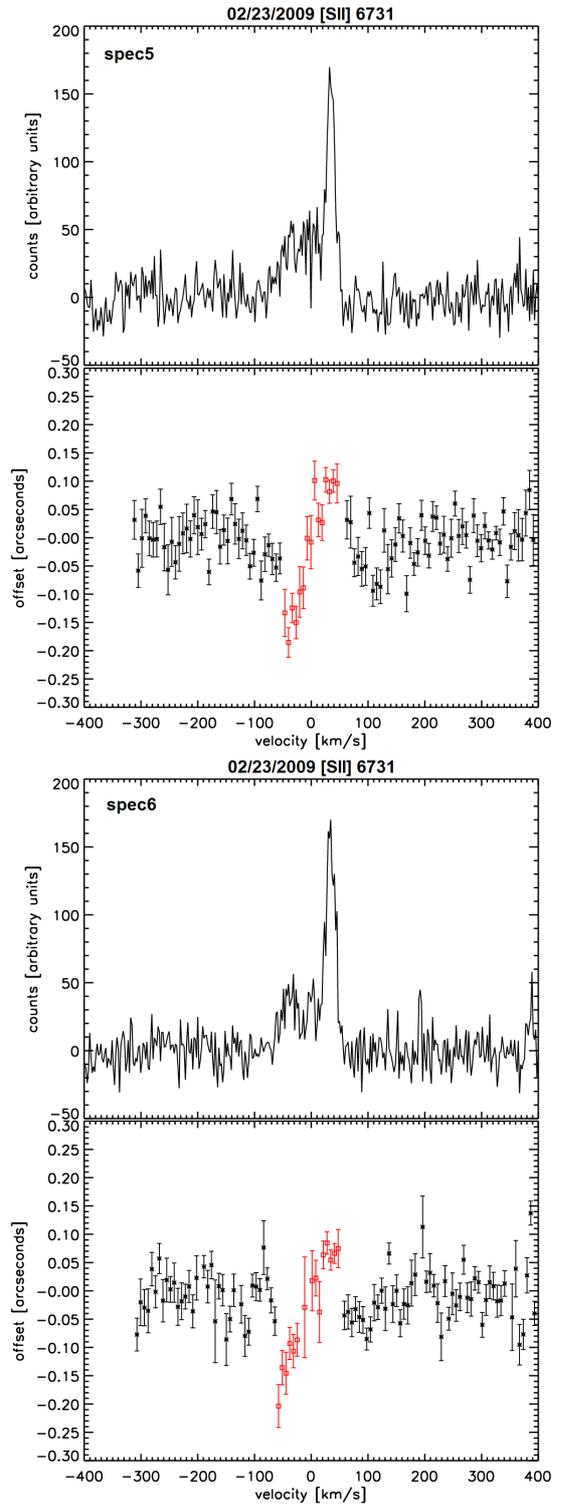}
\caption{
\label{fig:sa1}
{\bf Spectro-astrometric analysis of the [S\,II]$\lambda$6716 line} of
\iso217. \textit{Top panels}: 
Line profiles. Average of 8 pixel rows in the spatial direction centered on the continuum. 
The continuum has been subtracted. \textit{Bottom panels}: Spatial offset vs. radial velocity
of the continuum (black asterisks) and of the continuum subtracted FEL (red squares).
}
\end{figure}

\begin{figure}[t]
\centering
\includegraphics[width=.8\linewidth,clip]{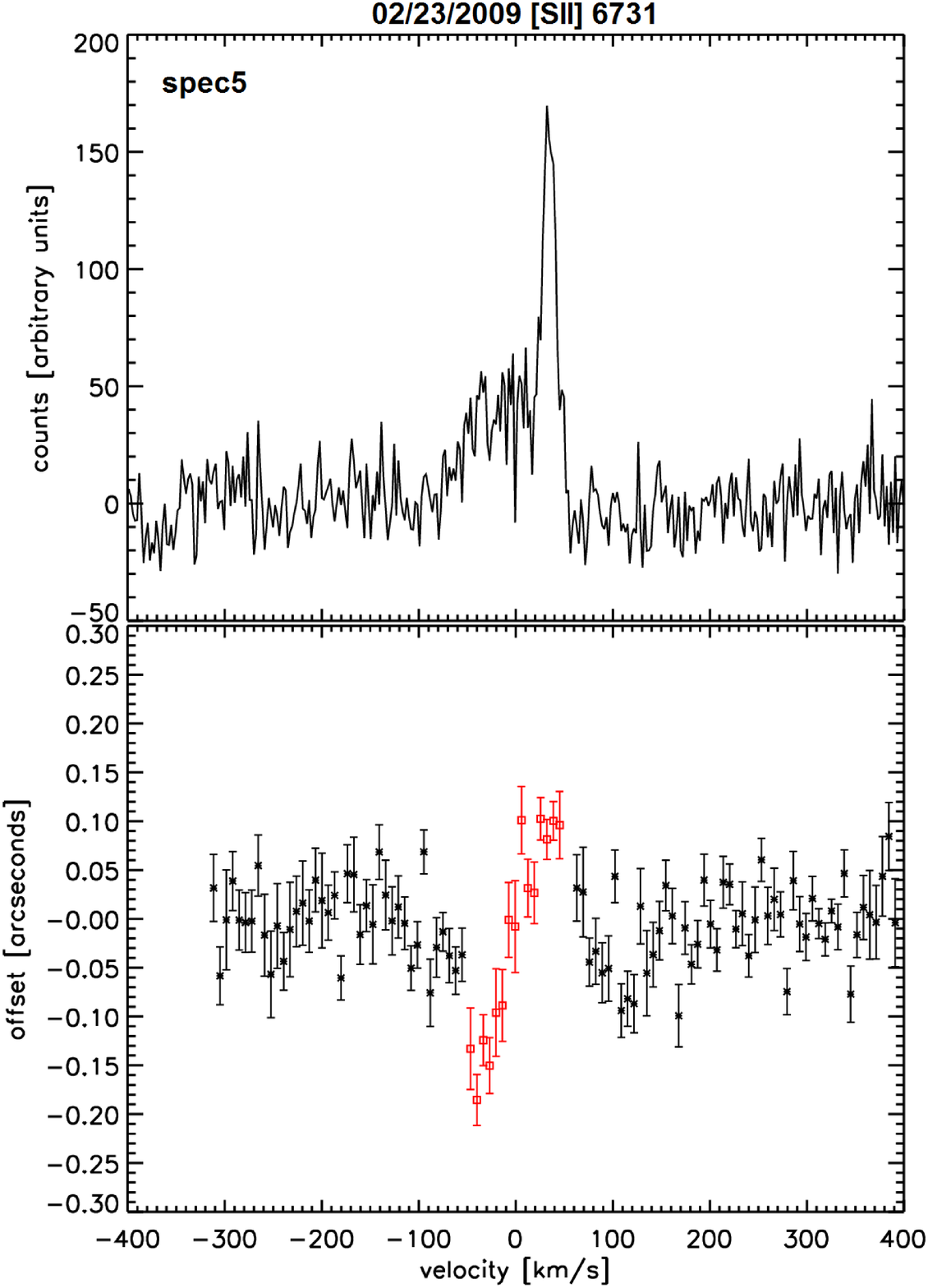}
\includegraphics[width=.8\linewidth,clip]{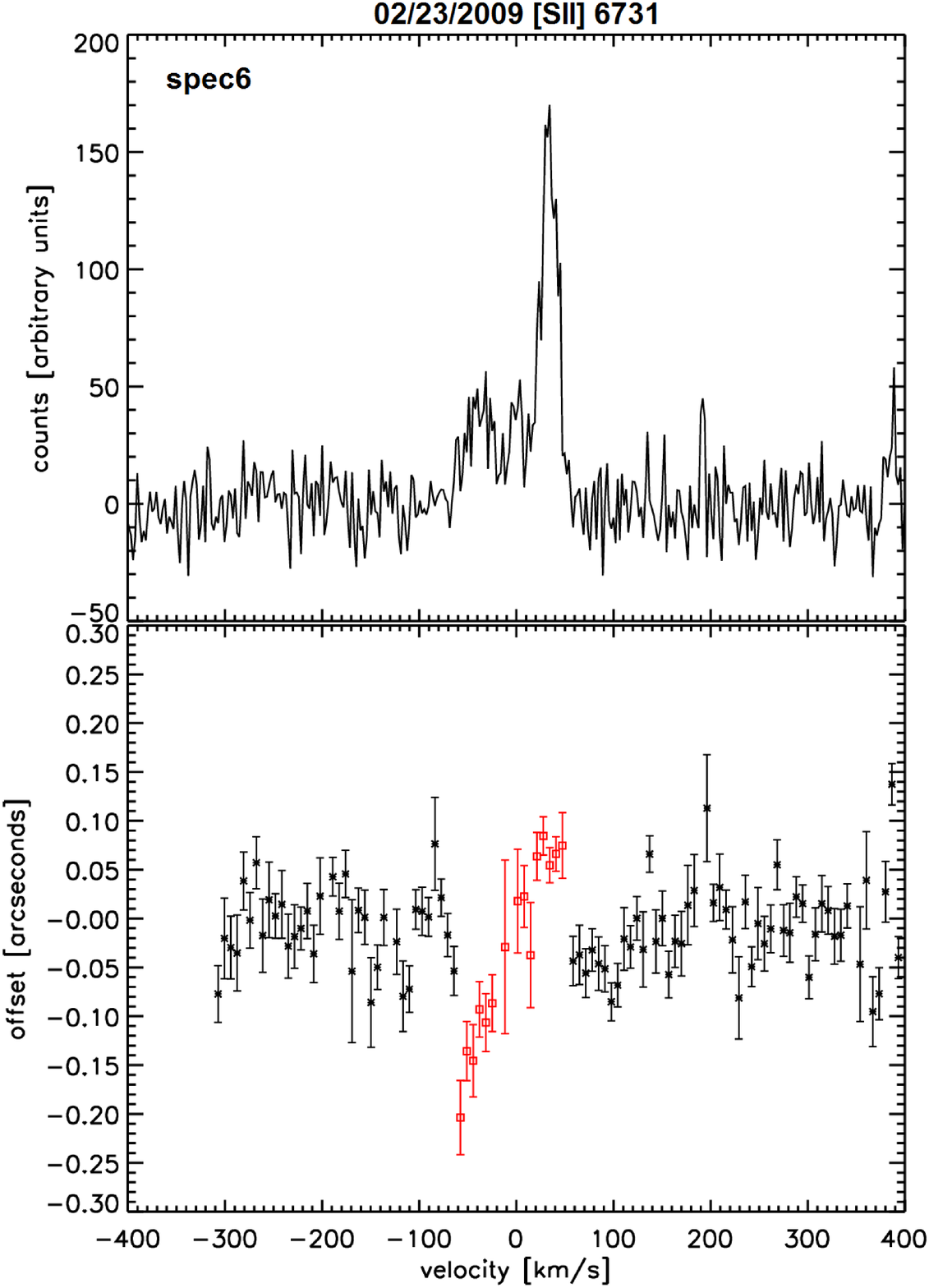}
\caption{
\label{fig:sa2}
{\bf Spectro-astrometric analysis of the [S\,II]$\lambda$6731 line} of \iso217. 
Otherwise same as Fig.\,\ref{fig:sa1}.
}
\end{figure}

{\em Forbidden [S\,II] emission.}
\iso217 has strong FELs of sulfur [S\,II] at 6716\,{\AA} and 6731\,{\AA}
(Figs.\, \ref{fig:sa1}-\ref{fig:sa2}) indicating an origin in a low density region.
We measured an EW of about -2\,\r{A} to -3\,\r{A} for [S\,II]$\lambda$6716
and of about -4\,\r{A} to -5\,\r{A} for [S\,II]$\lambda$6731, respectively (Table\,\ref{tab:ew}). 
It was shown by Whelan et al. (2009a) and confirmed
here (Sect.\,\ref{sect:results}) that these [S\,II] lines are produced in a bipolar outflow.
Both [S\,II] lines consist of two components, a blue-shifted one with an average peak velocity of -19\,km\,s$^{-1}$
and a red-shifted one with an average peak velocity of 34\,km\,s$^{-1}$ (Table\,\ref{tab:ew}).
The red component displays a sharp decline at its high velocity edge (between 40\,km\,s$^{-1}$ and 
50\,km\,s$^{-1}$), 
which is a typical outflow signature. 
The observed maximum radial velocities of the blue lobe of [S\,II] range from about -30\,km\,s$^{-1}$ 
to about -50\,km\,s$^{-1}$, though they are more difficult to estimate because of the weakness of this line wing.
Thus, we observed a velocity asymmetry in the peak emission with the red-shifted component being faster,
but this asymmetry is not so obvious in the maximum velocities of 
these lines (cf. also Sect.\,\ref{sect:results}).
It is noticeable that both outflow lobes of \iso217 are visible and that 
the red-shifted one is much stronger than the blue-shifted one.
This is uncommon for outflows of young stellar objects, 
because the circumstellar accretion disk usually obscures part of the red-shifted outflowing material.
To the best of our knowledge, this has only been observed for one other CTTS
(RW\,Aur, Hirth et al. 1994b).
The stronger red-shifted lobe that we see for the [S\,II] line of \iso217 
hints at an intrinsic asymmetry of the outflow of \iso217 
and at no or only little obscuration by the disk.

{\em Forbidden [Fe\,II] emission.}
\iso217 displays forbidden line emission in [Fe\,II] at 7155\,{\AA} (Fig.\,\ref{fig:feII}).
The measured EW is 1.6\,{\AA} and the line 
is emitted at relatively small blue-shifted velocities (-3\,km\,s$^{-1}$ at line peak).
Furthermore, there is tentative evidence of [Fe\,II] emission at 7172\,{\AA}.
Emission in [Fe\,II] has also been observed for jets of T~Tauri stars 
(e.g., Hartigan et al. 2004). 
We note that the [Fe\,II] line at 8617\,{\AA} is not covered by our observations
(because of a gap between CCD chips), which prevents us from
measuring the electron density in the densest regions of the outflow
based on the [Fe II] $\lambda$7155/$\lambda$8617 ratio
(cf. Hartigan et al. 2004).
Interestingly, we observe the [Fe\,II]$\lambda$7155\,{\AA} emission of \iso217 
solely from the blue-shifted component
and at significantly smaller velocities than the [S\,II] emission.
In the following, we show that the [Fe\,II] emission could originate 
from regions of higher temperatures than [S\,II]:
the ratio of the [S\,II] lines $\lambda$6716/$\lambda$6731 strongly depends on the
electron density and temperature. Considering 
the values observed for \iso217 (about 0.6, cf. Table\,\ref{tab:ew}), 
we derived electron densities of about 1--3$\times 10^3$\,cm$^{-3}$ 
for temperatures between 1000\,K and 10\,000\,K (Osterbrock 1989, p. 422). When restricting
the temperature range
to more reasonable values for a brown dwarf environment of 1000-3000\,K, 
we found a density range of 1-2 $\times 10^3$\,cm$^{-3}$. 
On the other hand, the [Fe\,II] line ratio $\lambda$7155/$\lambda$7172 is much more sensitive
to temperature 
than to electron density; for the considered densities it is practicably independent of the density.
We roughly estimated the observed [Fe\,II] $\lambda$7155/$\lambda$7172 line ratio of \iso217 
to about 4, which corresponds to a temperature in the range 2000-5000\,K
(NIST database, Ralchenko et al. 2011). Therefore, 
the differences observed between the [S\,II] and [Fe\,II] lines
could indicate that the [Fe\,II] emission is produced in a hotter
environment. 
A possible scenario is then that the [Fe\,II] emission
originates from the hot (and dense), inner regions of the outflow, 
where the bulk of the acceleration has not yet taken place,
and where occultation by the disk occurs more easily than
at greater radii.

{\em H$\alpha$ emission.}
In addition to our own data, we also measured EWs for  
the H$\alpha$ line in spectra from May 2007 taken by Whelan et al. (2009a). 
The EW of H$\alpha$ determined from these (pipeline-reduced and flux-calibrated) spectra is  
-137$\pm$7\,\r{A},
which is consistent with previous H$\alpha$ EW measurements for \iso217 (-137$\pm$7\,\r{A}, Luhman 2004;
between -70\,\r{A} and -230\,\r{A}, Muzerolle et al. 2005; Scholz \& Jayawardhana 2006).


\section{Spectro-astrometric analysis}
\label{sect:spectroastro}

We performed a spectro-astrometric analysis of the detected forbidden line emission
of [S\,II] in the two-dimensional (2-D) spectra of \iso217.
After completing a standard CCD reduction of the raw data (bias and flatfield correction and
cosmic-ray elimination), a row-by-row wavelength calibration of the 2-D 
spectra of individual echelle orders was done using the longslit package 
(fitcoords/transform) of IRAF. Finally, the sky was subtracted.
We note that the detected spectro-astrometric signatures of [S\,II]
are even visible in the unprocessed data, which demonstrates that  
the data reduction procedure does not introduce any artificial spatial offsets. 

We measured the spectro-astrometric signature in
the resulting 2-D spectra by Gaussian fitting the spatial
profile at each wavelength of both the FELs and the 
adjacent continuum, following e.g. Hirth et al. (1994a).
The spatial offset in the FEL was then computed relative to the continuum. 
In detail, the spatial position $y_C$ of the central source was first determined
by Gaussian fitting the continuum emission
in the spatial direction using spectral regions free of FELs. 
Secondly, the continuum emission was removed by fitting the continuum
on either side of the FEL row by row with a low-order polynomial
and subtracting these fits (cf. e.g., Davis et al. 2003).
This continuum subtraction plays an important role in revealing the 
weak FELs, as demonstrated e.g. by Hirth et al. (1994a). 
Finally, the spatial centroid position of the FEL $y'$ was measured in the 
continuum-subtracted spectrograms 
and the spatial offset $y = y' - y_C$ is computed as a function of the 
wavelength/velocity.
Velocities are given relative to the stellar rest-velocity of \iso217,
as determined in Sect.\,\ref{sect:obs2}.

A challenging aspect of the application of spectro-astrometry 
to high-resolution spectra of brown dwarfs and VLMS is the faintness of 
the emission lines. 
For the described spectro-astrometric analysis, the data 
were binned in the wavelength direction in order to increase the S/R,
as commonly done in the spectro-astrometry of brown dwarfs (e.g., Whelan et al. 2009a).
During the Gaussian fit procedure, the best-fit was found and its quality assessed 
by using the $\chi^{2}$-method.
In cases where the fit did not sufficiently represent the observed spectrum, 
these continuum data points were excluded from the plot.
It was found that the spectra taken in Feb 2009 have the highest S/R
and are best-suited for a spectro-astrometric analysis, whereas
the poorer quality of the spectra from Mar 2008 and Jan 2009 prevent us 
from performing a quantitative spectro-astrometry of these data. 

The UVES observations were originally optimized for radial velocity work, therefore,
the slit orientation was kept aligned with the direction of the atmospheric 
dispersion. 
As a consequence, the slit position angle (PA) changed relative 
to the sky during the observations:
each 2$\times$1500\,sec exposure samples a range of on-sky PAs
of about $\pm 7^{\circ}$, as given in Table\,\ref{tab:obslog}.
The mean slit PA of the spectra are between 130$^{\circ}$ and 180$^{\circ}$.
While a varying slit PA is not ideal for spectro-astrometry, for which one would like 
to keep the slit PA constant during an exposure,
the data allow nevertheless the detection of outflows of our target.

\begin{figure}[t]
\centering
\includegraphics[width=.8\linewidth,clip]{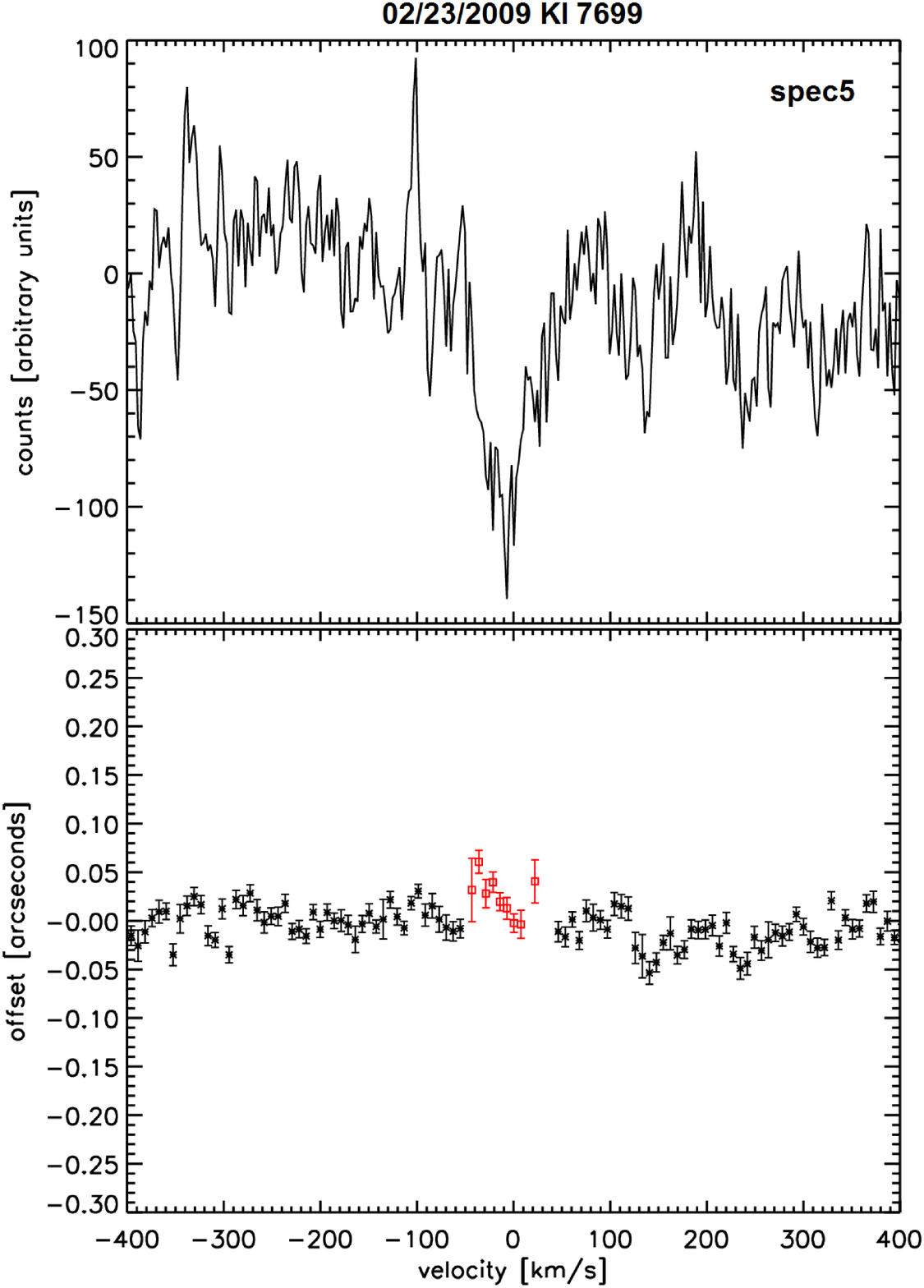}
\includegraphics[width=.8\linewidth,clip]{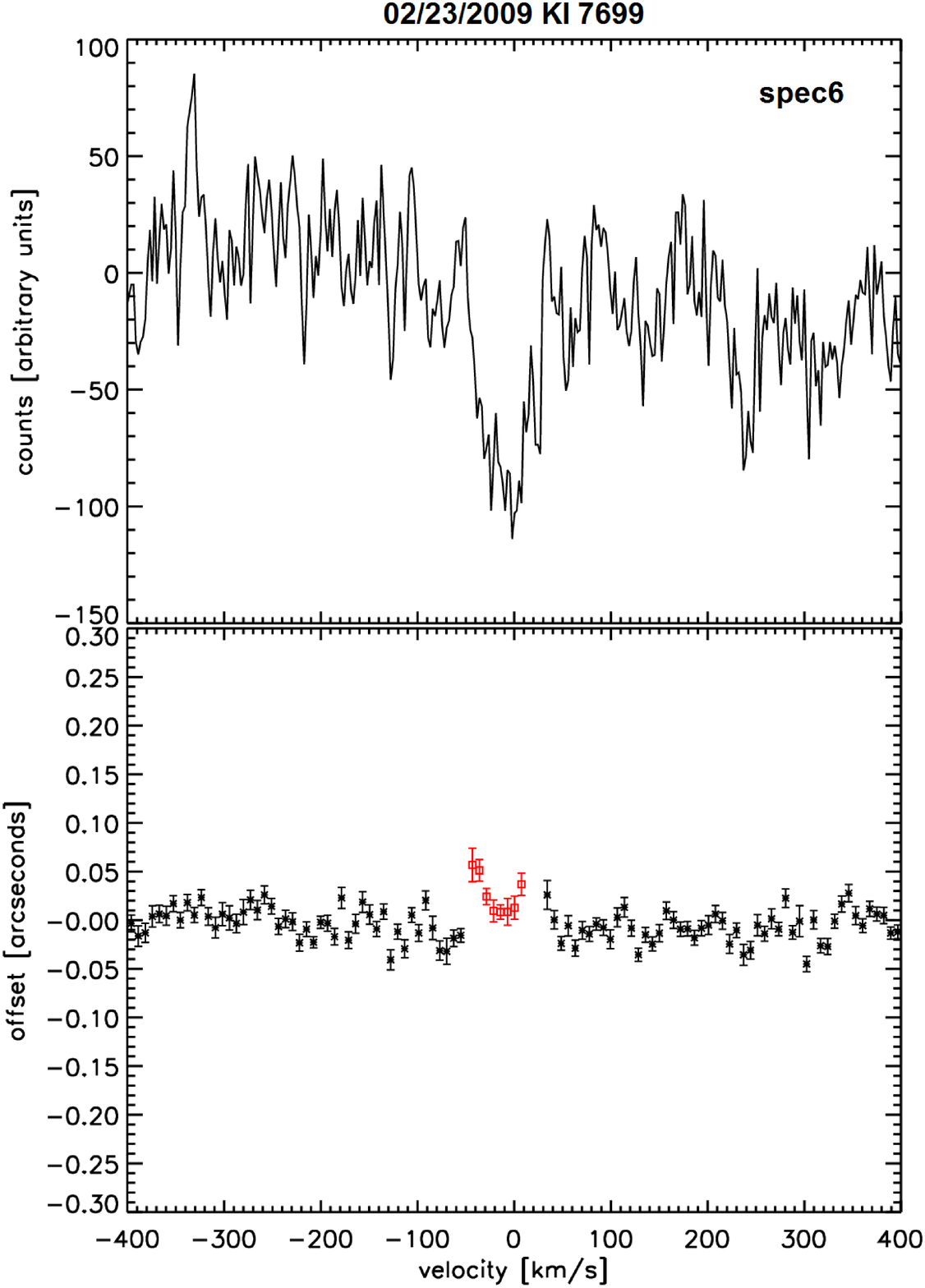}
\caption{
\label{fig:KI}
{\bf Spectro-astrometric analysis of photospheric line KI$\lambda$7699}
as a test for artifacts.
Otherwise same as Fig.\,\ref{fig:sa1}.
}
\end{figure}

To rule out spectro-astrometric artefacts, which can be caused for example by an asymmetric PSF 
(Brannigan et al. 2006),
we applied spectro-astrometry to a photospheric absorption line (KI\,$\lambda$7699)
and demonstrated that it has no spectro-astrometric signature
(spatial offset smaller than 50\,mas, see Fig.\,\ref{fig:KI}).

\section{Results of spectro-astrometry}
\label{sect:results}

\begin{table}[t]
\begin{center}
\caption{
\label{tab:spectroastro} 
FELs of \iso217.
}
\begin{tabular}{cclcr}
\hline
\hline
Slit PA [deg] & Line &  V [km\,s$^{-1}$] & Offset (mas) \\
\hline
\myrule
167.2\,$\pm$\,7.4 & [S\,II]$\lambda$6716 & -41\,/\,48 & -197\,/\,190 ~($\pm$42) \\  
(spec\,5)           & [S\,II]$\lambda$6731 & -40\,/\,45 & -167\,/\,102 ~($\pm$31) \\
\hline
\myrule
182.3\,$\pm$\,7.5 & [S\,II]$\lambda$6716 & -28\,/\,48 &  -82\,/\,146 ~($\pm$37) \\
(spec\,6)           & [S\,II]$\lambda$6731 & -54\,/\,47 & -175\,/\,78 ~($\pm$30)  \\ 
\hline
\end{tabular}
\tablefoot{
Given are the spatial offsets in milli-arcsec measured by spectro-astrometry
and the corresponding radial velocities. 
}
\end{center}
\end{table}

We have clearly detected the spectro-astrometric signature of a bipolar outflow in both [S\,II]
lines in our UVES spectra of \iso217.
For the quantitative spectro-astrometric analysis, we focused on the spectra taken in Feb 2009
because they have the highest S/R.
Figures\,\ref{fig:sa1}-\ref{fig:sa2} show in the top panels the line profiles of both the 
[S\,II]$\lambda\lambda$6716,6731 lines in these spectra
and in the bottom panels 
the measured spatial offsets of both the continuum and the FEL as a function of velocity.
The plotted errors in the spectro-astrometric plots are based on 1$\sigma$ errors in the Gaussian fit parameters.
Table\,\ref{tab:spectroastro} lists the maximum spatial offsets and corresponding velocities for both the blue 
and the red components of both [S\,II] lines.
These maximum spatial offsets were estimated from the weighted average of the two 
points with the largest offsets, to ensure that our measurement was less sensitive to outliers. 
Offset errors in Table\,\ref{tab:spectroastro} are based on the standard deviation in the continuum points.
For an overview and to constrain the outflow PA (see below), we plot the maximum spatial offsets 
in Fig.\,\ref{fig:pa} as a function of the slit PA.

Our spectro-astrometric analysis of the detected FELs of [S\,II] demonstrated
that they originate from spatially offset positions on either side of
the continuum source of up to $\pm$190\,mas 
(about $\pm$30\,AU at the distance of Cha\,I)
at a velocity of up to $\pm$40-50\,km\,s$^{-1}$. 

We found the [S\,II]$\lambda$6716 emission to be 
spatially more extended than the [S\,II]$\lambda$6731 emission
in all but one case (Table\,\ref{tab:spectroastro}, Fig.\,\ref{fig:pa}),
which is consistent with the [S\,II]$\lambda$6716 line
tracing lower densities than the [S\,II]$\lambda$6731 line.

The asymmetry seen for the two lobes in the [S\,II] line profiles,
where the red lobe is much stronger and also faster 
(top panels in Fig.\,\ref{fig:sa1}-\ref{fig:sa2}, Sect.\,\ref{sect:lines}), 
is a remarkable feature of the \iso217 outflow.
In accordance with this, we observed 
in the spectro-astrometry of these lines a tendency for the blue lobe to be slower and more extended:
(i) The measured outflow velocities range between -30\,km\,s$^{-1}$ and -50\,km\,s$^{-1}$ for the blue lobe, 
and are about 50\,km\,s$^{-1}$ for the red lobe (Table\,\ref{tab:spectroastro}). 
(ii) The spatial offsets measured for the blue lobe are larger than those for the red lobe 
in all but one case (Table\,\ref{tab:spectroastro}, Fig.\,\ref{fig:pa}). 
The exception is the [S\,II]$\lambda$6716 line of spec\,6, for which the offset measurement might be 
affected by an outlying data point (Fig.\,\ref{fig:sa1}). 

The spectra were taken at mean slit PAs of 167$^{\circ}\,\pm\,7^{\circ}$ and 182$^{\circ}\,\pm\,8^{\circ}$ 
(cf. Sect.\,\ref{sect:spectroastro}),
i.e. not very far from the value determined for the outflow PA of \iso217 from the 
2007 observations of Whelan et al. (2009a,
193-200$^{\circ}$ for [S\,II]$\lambda \lambda$6716,6731).
As illustrated in Fig.\,\ref{fig:pa},
we found for an increasing mean slit PA from $167^{\circ}$ to $182^{\circ}$,
either a slight decrease ([S\,II]$\lambda$6716 line) in or a constant level ([S\,II]$\lambda$6731)
of the observed spatial extension of the outflow.
Hence, the data are consistent with an outflow angle that is closer to $170^{\circ}$ than $180^{\circ}$.
However, the projection of a homogenous outflow with an angle of $200^{\circ}$ and an extension of 190\,mas 
onto a slit with a PA of 167$^{\circ}$ would decrease the observable extension 
by $\sim$30\,mas, which is
on the order of the precision of the spectro-astrometry (30-40\,mas) performed both 
here and by Whelan et al. (2009a).
The measurement of a different outflow angle for a slightly different slit PA could also be explained by 
an outflow that has a wide opening angle, causing an almost constant spatial offset for a large range of PAs 
and/or by an outflow that has a common knot structure
(e.g., Hirth et al. 1994b) and for which
the measured spatial offset depends on whether a knot is detected for a certain slit PA 
or for example an outside edge of such a knot.

\begin{figure}[t]
\centering
\includegraphics[width=.8\linewidth,clip]{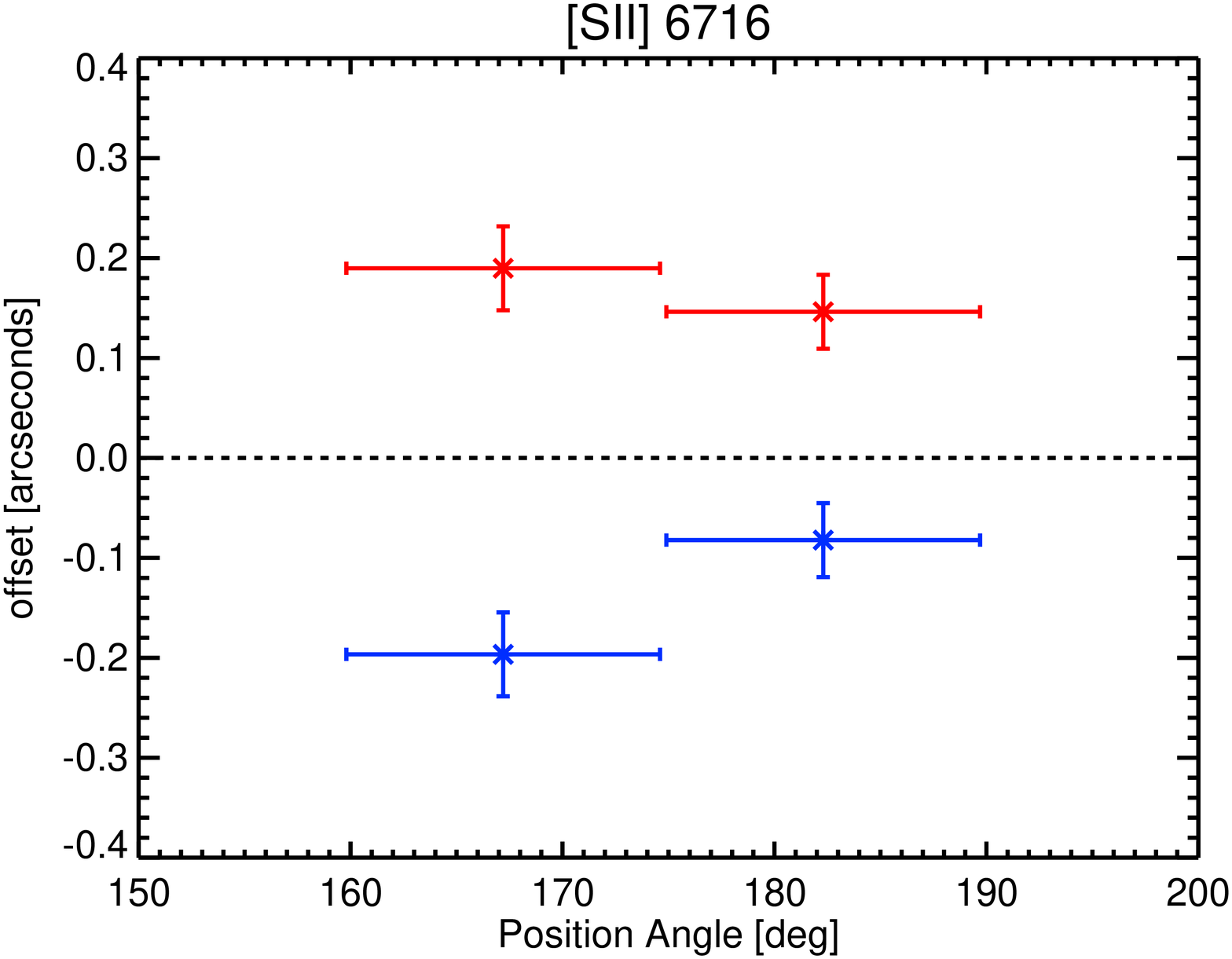}
\includegraphics[width=.8\linewidth,clip]{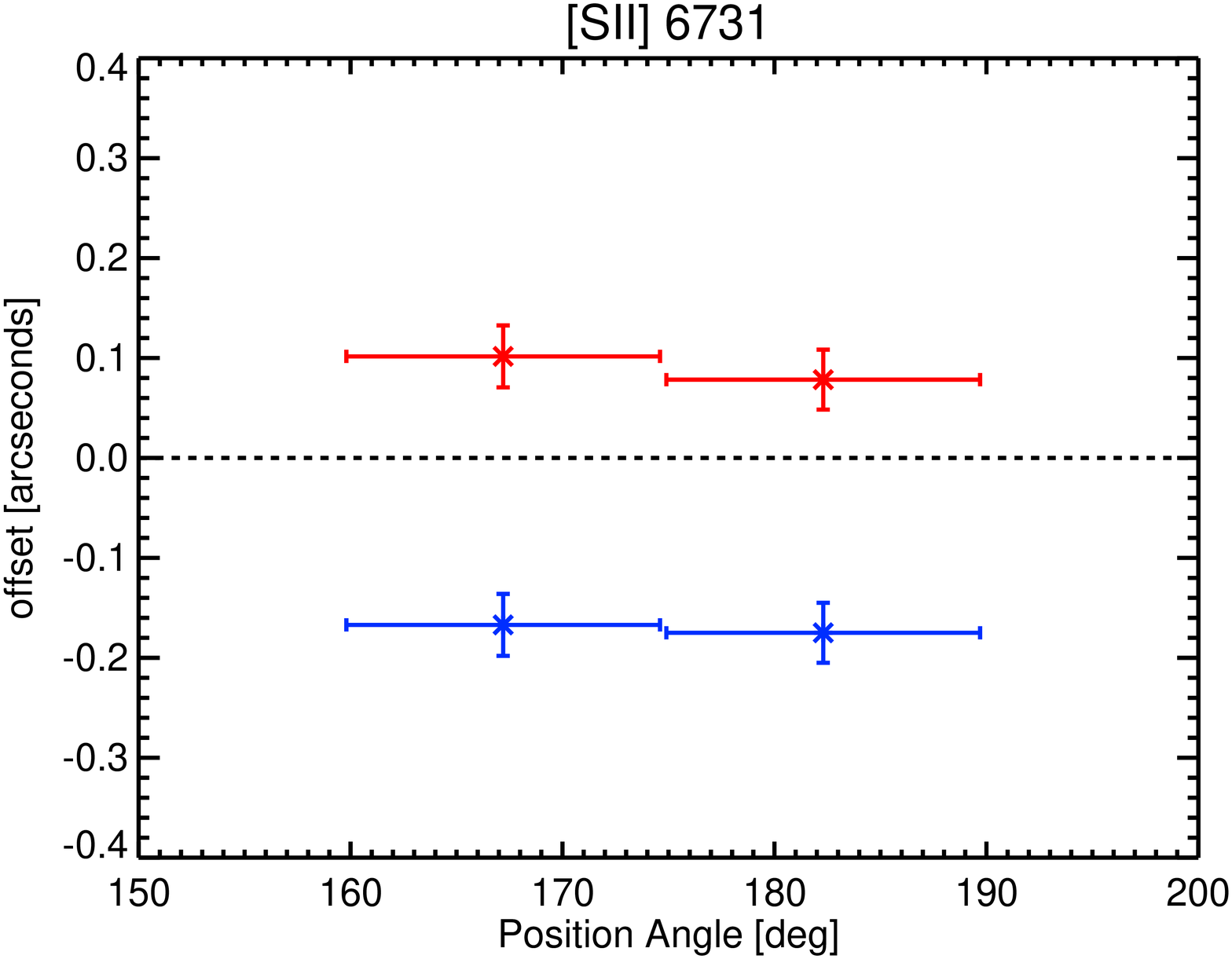}
\caption{
\label{fig:pa}
{\bf Spatial offsets of FELs of \iso217 as a function of the slit PA} 
for [S\,II]$\lambda$6716 (top panel) and [S\,II]$\lambda$6731 (bottom panel). 
}
\end{figure}


\section{Discussion and conclusions}
\label{sect:concl}

We have proven that the very young brown dwarf candidate \iso217 (M6.25) is 
driving a bipolar outflow with a stronger and faster red-shifted component based on spectro-astrometry 
of [S\,II] lines in UVES/VLT spectra taken in 2009.
\iso 217 is only one of a handful of brown dwarfs and VLMS
(M5-M8) for which an outflow has been detected.
We have demonstrated that the forbidden [S\,II] emission in \iso217 originates from spatially offset 
positions straddling the central source
by up to $\pm$190\,mas 
(about $\pm$30\,AU at the distance of Cha\,I) at a velocity of up to $\pm$40-50\,km\,s$^{-1}$. 
There is also evidence of forbidden [Fe\,II] emission of the blue-shifted component of this outflow.
Given the excitation potential of this line, its relatively small velocity ($\sim$3\,km\,s$^{-1}$),
and the observation of solely the blue component, 
we propose as a potential origin of the [Fe\,II] emission
the dense innermost regions, where the outflow has not yet been accelerated.

We have detected a velocity asymmetry between the two outflow lobes.
This can be seen in the [S\,II] line profiles, where 
the blue-shifted component has an average peak velocity of -19\,km\,s$^{-1}$
and the red-shifted one of 34\,km\,s$^{-1}$. 
Furthermore, the outflow velocities measured using spectro-astrometry 
tend to be smaller for the blue lobe (between -30\,km\,s$^{-1}$ and -50\,km\,s$^{-1}$) than
for the red lobe (about 50\,km\,s$^{-1}$).
There is tentative evidence that the two lobes are also spatially asymmetric, 
with the blue component being more extended.

Intrinsic changes in an outflow can occur on timescales of its travel time,
which is shorter than three years for \iso217 given the observed outflow velocity and spatial extension.
We have investigated the possible differences of the outflow properties 
that have been inferred here based on spectra from Feb 2009
and those determined in the discovery spectra from May 2007 (Whelan et al. 2009a).
This is described in detail in the next few paragraphs. Our main results are that
we have found that the basic features of the \iso217 outflow (spatial extension, velocities, and outflow PA) 
are of similar order in 2007 and 2009, and that
the velocity asymmetry between both lobes seem to have decreased slightly in this time period.
In addition, we have demonstrated that the strong velocity asymmetry between both lobes of a factor of two found in 2007
might be smaller than originally anticipated when using a more realistic stellar rest-velocity.

The detailed comparison of the outflow properties in 2007 to those in 2009 is described in the following.
The line profiles of both [S\,II] lines have a very similar shape from 2007 to 2009.
The noteable differences in the line profiles are 
slightly less asymmetric peak velocities in 2009 
(the difference between blue and red peak of [S\,II]$\lambda$6731 in Feb 2009 is 17\,\,km\,s$^{-1}$ compared to 24\,\,km\,s$^{-1}$ in 2007
at PA=0$^{\circ}$)
and a slight decrease in the line strength (the EW of [S\,II]$\lambda$6731 is -4\,\r{A} in 2009 
compared to -6\,\r{A} in 2007).
Furthermore, 
the peak velocities of the [S\,II] lines (both lobes) appear to be generally shifted by a few km\,s$^{-1}$ 
towards the blue from 2007 to 2008/2009. This implies that the velocity of the outflow varied during this time period.
We note that the velocities considered in this comparison of the peak velocities
were consistently measured using the same stellar rest-velocity of $V_{0}=17.2$\,km\,s$^{-1}$,
which was derived here for the 2008/2009 spectra.

The spatial extension of the outflow measured by spectro-astrometry in the [S\,II] lines 
in 2009 (80-200\,$\pm$\,35\,mas) at slit position angles of
167$^{\circ}$\,$\pm$\,7$^{\circ}$ and 182$^{\circ}$\,$\pm$\,8$^{\circ}$ 
is to a large degree consistent 
with that measured in 2007 (180\,$\pm$\,34\,mas) at a slit position angle of $0^{\circ}$ (corresponding to $180^{\circ}$,
Whelan et al. 2009a). 
In addition, the observed outflow velocities are of similar order, although, we have found that they are 
less asymmetric than in 2007:
in 2009, the velocity at the maximum spatial offset is about 50\,km\,s$^{-1}$ for the red lobe and 
between -30\,km\,s$^{-1}$ and -50\,km\,s$^{-1}$ for the blue lobe.
The velocities measured by Whelan et al. (2009a) for the 2007 spectra, however, seem to be more asymmetric, 
with the red-shifted component (40\,km\,s$^{-1}$)
having a velocity approximately twice as large as the blue-shifted one 
(-20\,km\,s$^{-1}$).
This might be attributed partly to the use of a potentially inaccurate stellar rest-velocity:
Whelan et al. (2009a) adopt a stellar rest-velocity of 12.6\,km\,s$^{-1}$, 
which is an estimated mean value for T~Tauri stars in Cha\,I. 
However, this is almost 5\,km\,s$^{-1}$ smaller than the stellar rest-velocity we determined here for 
and applied to \iso217 in 2008/2009.

We found that for an increasing slit PA from $167^{\circ}$ to $182^{\circ}$,
the observed spatial extension of the outflow
for both lobes in both [S\,II] lines
either slightly decreases or remains at a constant level.
Hence, our data imply that the outflow angle is closer to $170^{\circ}$ than $180^{\circ}$, i.e. that it is
slightly smaller than but within the errors still consistent
with the value measured based on two orthogonal spectra  
(193--200$^{\circ}$ in the [S\,II] lines) by Whelan et al. (2009a).

The outflow activity of young stellar objects is intrinsically tied to the accretion disk.
To gain a comprehensive understanding of the \iso217 outflow and disk system,
we have determined the disk properties of \iso217 by performing a radiative transfer modeling of its SED.
The accretion disk surrounding \iso217 has a total mass of 4$\times 10^{-6} M_\odot$,
a flared geometry, and is viewed under an inclination angle of about 45$^{\circ}$
according to our model that most accurately fits the observed SED 
from 0.66 - 24\,$\mu$m and that is also in
very good agreement with Herschel/PACS observations at 70\,$\mu$m (Harvey et al. 2012b).
We have shown that a disk inclination significantly exceeding 45$^{\circ}$, as 
previously suggested based on H$\alpha$ modeling (65$^{\circ}$, Muzerolle et al. 2005)
and the visibility of both lobes of the outflow (Whelan et al. 2009a),
is inconsistent with the SED data.
The disk mass of only about one Earth mass is very low for a CTTS disk,
but fully consistent with that of other brown dwarfs and VLMS 
(10$^{-5} - 10^{-6} M_\odot$, e.g., Harvey et al. 2012a).
However, it is strikingly low compared to the estimated accretion
($1\times10^{-10}\,\mperyr$, Muzerolle et al. 2005) 
and mass-loss rate (2-3$\times10^{-10}\,\mperyr$, Whelan et al. 2009a) of \iso217. 
Possible explanations for this discrepancy between the dust-inferred disk mass 
and the gas-inferred accretion and outflow mass rates
are one or both of the following:
(i) The disk mass is higher than derived from the model because of 
a population of large particles ($>$100 $\mu$m) that remain undetected in the available SED data ($\leq$70 $\mu$m).
(ii) The mass accretion and outflow rates are on average lower than the determined values, 
which is plausible in the case of the accretion rate given the signs of variable accretion of \iso217.
In the case of undetected strong grain growth, the disk could also be in a more evolved phase 
than inferred from its flared appearance.

Apart from accretion through the disk and a bipolar outflow, 
there is also evidence of material falling onto \iso217 at velocities of up to $130\,$km\,s$^{-1}$
(from Ca\,II IR emission with a red-shifted large-velocity tail, this work) and of
a wind expanding at a velocity of up to 30\,km\,s$^{-1}$ 
(Scholz \& Jayawardhana 2006; Whelan et al. 2009a). 

To summarize, \iso217 is a very young M6.25 type object at the substellar limit,  
that is surrounded by a flared, intermediately inclined accretion disk with possibly strong 
grain growth. It is driving a bipolar outflow detected in FELs of [S\,II] with a $\pm$30\,AU spatial extension
and velocities of up to $\pm$40-50\,km\,s$^{-1}$. The outflow is intrinsically asymmetric with a 
stronger and slightly faster red-shifted component. 
We have found this velocity asymmetry to be variable on timescales of the outflow travel time. 
The predominance of the red outflow component detected in \iso217 is very rarely seen for CTTS (only one other 
case is known, Hirth et al. 1994b) and it implies that 
the disk does not appear to obscure the red-shifted outflow lobe
despite its intermediate inclination angle.

\begin{acknowledgements}
We would like to acknowledge the excellent work of the ESO staff at Paranal, 
who took the data analyzed here in service mode. 
We thank E. Whelan for valuable discussions on spectro-astrometric analysis 
and for providing UVES spectra from her program.
We also thank P. Harvey, F. Menard, S. Wolf and the rest of the Herschel low-mass disk team
for providing the PACS flux data for \iso217 prior to publications.
We are grateful for very helpful comments from the referee, T. Ray, which
allowed us to improve the paper.
Furthermore, we thank 
I. Pascucci, R. Mundt, R. van Boekel, and C. Dullemond for interesting discussions 
on varies topics of this paper, as well as
R. Smiljanic from the ESO USD for information on position angles.
Part of this work was funded by the ESF in Baden-W\"urttemberg.
A.S.A. acknowledges support by the Spanish "Ram\'{o}n y Cajal" program.
\end{acknowledgements}

\end{document}